\documentstyle[twoside,epsfig]{article}

\catcode`\@=11
\long\def\@makefntext#1{
\protect\noindent \hbox to 3.2pt {\hskip-.9pt  
$^{{\eightrm\@thefnmark}}$\hfil}#1\hfill}		

\def\@makefnmark{\hbox to 0pt{$^{\@thefnmark}$\hss}}	
	
\def\ps@myheadings{\let\@mkboth\@gobbletwo
\def\@oddhead{\hbox{}
\rightmark\hfil\eightrm\thepage}   
\def\@oddfoot{}\def\@evenhead{\eightrm\thepage\hfil
\leftmark\hbox{}}\def\@evenfoot{}
\def\sectionmark##1{}\def\subsectionmark##1{}}

\oddsidemargin=\evensidemargin
\newcounter{sectionc}\newcounter{subsectionc}\newcounter{subsubsectionc}
\renewcommand{\section}[1] {\vspace{12pt}\addtocounter{sectionc}{1} 
\setcounter{subsectionc}{0}\setcounter{subsubsectionc}{0}\noindent 
	{\tenbf\thesectionc. #1}\par\vspace{5pt}}
\renewcommand{\subsection}[1] {\vspace{12pt}\addtocounter{subsectionc}{1} 
	\setcounter{subsubsectionc}{0}\noindent 
	{\bf\thesectionc.\thesubsectionc. {\kern1pt \bfit #1}}\par\vspace{5pt}}
\renewcommand{\subsubsection}[1] {\vspace{12pt}\addtocounter{subsubsectionc}{1}
	\noindent{\tenrm\thesectionc.\thesubsectionc.\thesubsubsectionc.
	{\kern1pt \tenit #1}}\par\vspace{5pt}}
\newcommand{\nonumsection}[1] {\vspace{12pt}\noindent{\tenbf #1}
	\par\vspace{5pt}}

\newcounter{appendixc}
\newcounter{subappendixc}[appendixc]
\newcounter{subsubappendixc}[subappendixc]
\renewcommand{\thesubappendixc}{\Alph{appendixc}.\arabic{subappendixc}}
\renewcommand{\thesubsubappendixc}
	{\Alph{appendixc}.\arabic{subappendixc}.\arabic{subsubappendixc}}

\renewcommand{\appendix}[1] {\vspace{12pt}
        \refstepcounter{appendixc}
        \setcounter{figure}{0}
        \setcounter{table}{0}
        \setcounter{lemma}{0}
        \setcounter{theorem}{0}
        \setcounter{corollary}{0}
        \setcounter{definition}{0}
        \setcounter{equation}{0}
        \renewcommand{\thefigure}{\Alph{appendixc}.\arabic{figure}}
        \renewcommand{\thetable}{\Alph{appendixc}.\arabic{table}}
        \renewcommand{\theappendixc}{\Alph{appendixc}}
        \renewcommand{\thelemma}{\Alph{appendixc}.\arabic{lemma}}
        \renewcommand{\thetheorem}{\Alph{appendixc}.\arabic{theorem}}
        \renewcommand{\thedefinition}{\Alph{appendixc}.\arabic{definition}}
        \renewcommand{\thecorollary}{\Alph{appendixc}.\arabic{corollary}}
        \renewcommand{\theequation}{\Alph{appendixc}.\arabic{equation}}
        \noindent{\tenbf Appendix \theappendixc #1}\par\vspace{5pt}}
\newcommand{\subappendix}[1] {\vspace{12pt}
        \refstepcounter{subappendixc}
        \noindent{\bf Appendix \thesubappendixc. {\kern1pt \bfit #1}}
	\par\vspace{5pt}}
\newcommand{\subsubappendix}[1] {\vspace{12pt}
        \refstepcounter{subsubappendixc}
        \noindent{\rm Appendix \thesubsubappendixc. {\kern1pt \tenit #1}}
	\par\vspace{5pt}}

\newcommand{\textlineskip}{\baselineskip=13pt}
\newcommand{\smalllineskip}{\baselineskip=10pt}

\def\eightcirc{
\begin{picture}(0,0)
\put(4.4,1.8){\circle{6.5}}
\end{picture}}
\def\eightcopyright{\eightcirc\kern2.7pt\hbox{\eightrm c}} 

\newcommand{\copyrightheading}[1]
	{\vspace*{-2.5cm}\smalllineskip{\flushleft
	{\footnotesize International Journal of Modern Physics A, #1}\\
	{\footnotesize $\eightcopyright$\, World Scientific Publishing
	 Company}\\
	 }}

\def\abstracts#1#2#3{{
	\centering{\begin{minipage}{4.5in}\baselineskip=10pt\footnotesize
	\parindent=0pt #1\par 
	\parindent=15pt #2\par
	\parindent=15pt #3
	\end{minipage}}\par}} 

\renewenvironment{thebibliography}[1]
	{\frenchspacing
	 \ninerm\baselineskip=11pt
	 \begin{list}{\arabic{enumi}.}
	{\usecounter{enumi}\setlength{\parsep}{0pt}
	 \setlength{\leftmargin 12.7pt}{\rightmargin 0pt} 
	 \setlength{\itemsep}{0pt} \settowidth
	{\labelwidth}{#1.}\sloppy}}{\end{list}}

\newcounter{itemlistc}
\newcounter{romanlistc}
\newcounter{alphlistc}
\newcounter{arabiclistc}

\newcommand{\fcaption}[1]{
        \refstepcounter{figure}
        \setbox\@tempboxa = \hbox{\footnotesize Fig.~\thefigure. #1}
        \ifdim \wd\@tempboxa > 5in
           {\begin{center}
        \parbox{5in}{\footnotesize\smalllineskip Fig.~\thefigure. #1}
            \end{center}}
        \else
             {\begin{center}
             {\footnotesize Fig.~\thefigure. #1}
              \end{center}}
        \fi}

\newcommand{\tcaption}[1]{
        \refstepcounter{table}
        \setbox\@tempboxa = \hbox{\footnotesize Table~\thetable. #1}
        \ifdim \wd\@tempboxa > 5in
           {\begin{center}
        \parbox{5in}{\footnotesize\smalllineskip Table~\thetable. #1}
            \end{center}}
        \else
             {\begin{center}
             {\footnotesize Table~\thetable. #1}
              \end{center}}
        \fi}

\def\@citex[#1]#2{\if@filesw\immediate\write\@auxout
	{\string\citation{#2}}\fi
\def\@citea{}\@cite{\@for\@citeb:=#2\do
	{\@citea\def\@citea{,}\@ifundefined
	{b@\@citeb}{{\bf ?}\@warning
	{Citation `\@citeb' on page \thepage \space undefined}}
	{\csname b@\@citeb\endcsname}}}{#1}}

\newif\if@cghi
\def\cite{\@cghitrue\@ifnextchar [{\@tempswatrue
	\@citex}{\@tempswafalse\@citex[]}}
\def\citelow{\@cghifalse\@ifnextchar [{\@tempswatrue
	\@citex}{\@tempswafalse\@citex[]}}
\def\@cite#1#2{{$\null^{#1}$\if@tempswa\typeout
	{IJCGA warning: optional citation argument 
	ignored: `#2'} \fi}}

\def\pmb#1{\setbox0=\hbox{#1}
	\kern-.025em\copy0\kern-\wd0
	\kern.05em\copy0\kern-\wd0
	\kern-.025em\raise.0433em\box0}

\def\fnt#1#2{\footnotetext{\kern-.3em
	{$^{\mbox{\scriptsize #1}}$}{#2}}}

\def\fpage#1{\begingroup
\voffset=.3in
\thispagestyle{empty}\begin{table}[b]\centerline{\footnotesize #1}
	\end{table}\endgroup}

\def\runninghead#1#2{\pagestyle{myheadings}
\markboth{{\protect\footnotesize\it{\quad #1}}\hfill}
{\hfill{\protect\footnotesize\it{#2\quad}}}}
\font\tenrm=cmr10
\font\tenit=cmti10 
\font\tenbf=cmbx10
\font\bfit=cmbxti10 at 10pt
\font\ninerm=cmr9

\font\eightrm=cmr8

\def\qed{\hbox{${\vcenter{\vbox{			
   \hrule height 0.4pt\hbox{\vrule width 0.4pt height 6pt
   \kern5pt\vrule width 0.4pt}\hrule height 0.4pt}}}$}}

\begin{document}

\runninghead{A Novel Optical Package for ATLAS Pixel Detector}
            {A Novel Optical Package for ATLAS Pixel Detector}

\normalsize\textlineskip
\thispagestyle{empty}
\setcounter{page}{1}

\copyrightheading{}			

\vspace*{0.88truein}

\fpage{1}
\centerline{\bf A Novel Optical Package for ATLAS Pixel Detector}
\vspace*{0.37truein}
\centerline{\footnotesize K.K.~Gan}
\vspace*{0.015truein}
\centerline{\footnotesize\it Department of Physics,
        The Ohio State University,
        Columbus, OH 43210,
        U.S.A.}

\vspace*{0.21truein}
\abstracts{An optical package of novel design has been developed for the ATLAS pixel detector.
The package contains two VCSELs and one PIN diode to transmit and receive optical signals.
The design is based on a simple connector-type concept and is made of
radiation-hard material.
Several packages have been fabricated and show promising results.}{}{}


\vspace*{1pt}\textlineskip	
\section{Introduction}	
\vspace*{-0.5pt}
\noindent
The ATLAS pixel detector\cite{pixel} consists of three barrel layers and
three forward and backward disks.
The detector covers the pseudo-rapidity region $| \eta | < 2.5$ and
provides at least three space point measurements.
The signal from the digitalization electronics of the pixel detector
is converted into an optical signal using a Vertical Cavity Surface
Emitting Laser (VCSEL)\cite{Mitel} and is transmitted to the counting
room via a fibre.
The 40-MHz beam crossing clock, encoded with the command signal, from
the counting room, is transmitted to a PIN diode\cite{Centronics} via a fibre.
The VSCEL and PIN couple to the two fibres inside an optical package
located in close proximity to the pixel detector.
In this paper, we describe the performance of a more complicated package
which contains two VSCELs and one PIN diode.
The simpler version with no redundancy is adopted by the ATLAS pixel collaboration
due to the severe space constraints around the pixel detector region.

\section{Optical Package Design}
\vspace*{-0.5pt}
\noindent
The main technical challenge in the fabrication of the package is
the tight alignment tolerance of the VCSEL with respect to the fibre:

\hspace{0.1in} $\bullet$ 50 $\mu$m in $z$ (along the fibre)

\hspace{0.1in} $\bullet$ 25 $\mu$m in $r$ (transverse to the fibre)

\noindent The requirements can be satisfied either by passive or
active alignment.
For the former, parts must be fabricated or placed with high precision
($\le 10~\mu$m) so that total precision is still within the tolerance.

There are currently three designs for the optical package:
Marconi, Ohio State University (OSU), and Academia Sinica (Taiwan).
Both Marconi and OSU use passive alignment while Taiwan uses active alignment.
The Marconi design consists of several pieces and uses
aluminized mirrors to reflect the light by 45$^0$.
The Taiwan design consists of three pieces and
replaces the mirrors by cleaving the fibers at 45$^0$.
Each fiber is actively aligned and glued permanently to the package.
The package is of low cost but the permanent attachment of the fibers
to the package presents a technical challenge to the assembly of
the pixel detector.

\begin{figure}[htbp]
\centerline{\psfig{figure=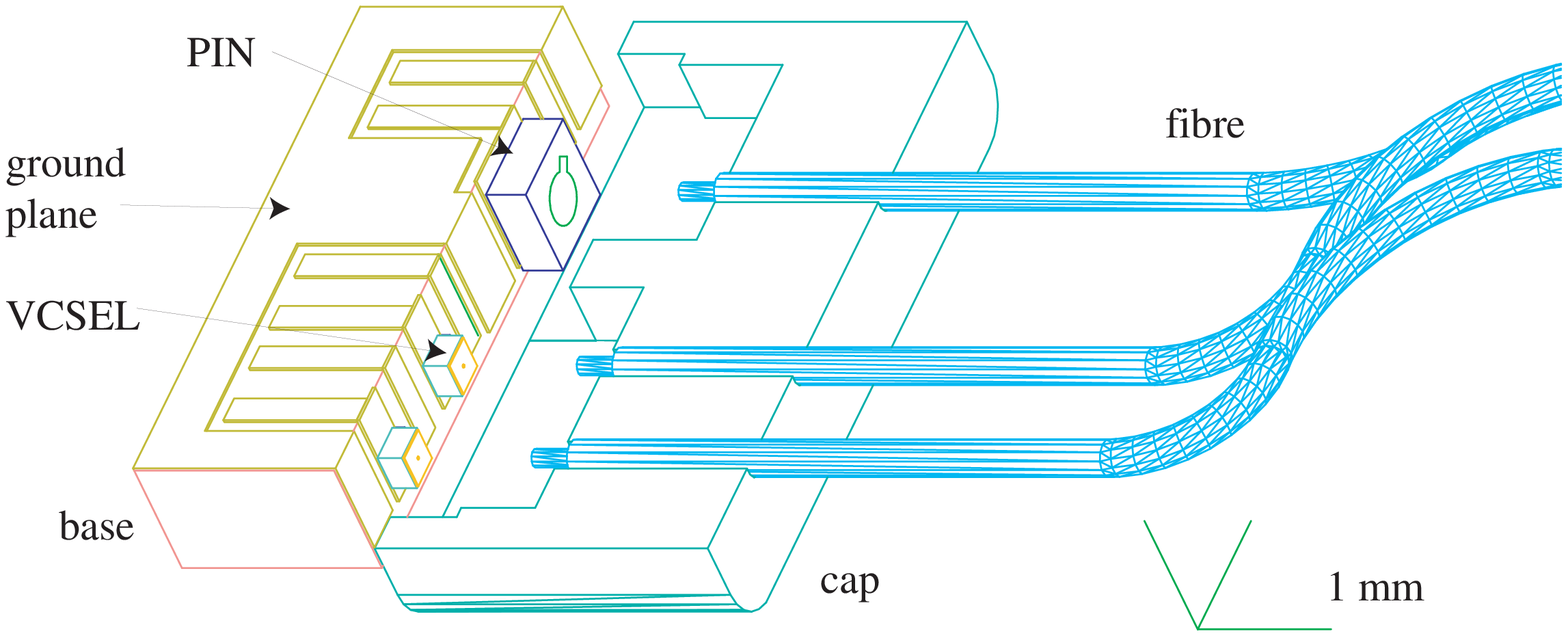,width=2.5in,angle=0}}
\vspace{-.3in}
\fcaption{
A cut out view of the optical package showing the two VCSELs and
one PIN diode mounted on gold pads which are connected to the gold traces.}
\label{cap_base_fibre}
\end{figure}

The OSU design uses a connector concept: a cap with three holes
for the fibres and a base with deposited wire bonding traces
and pads for PIN and VCSEL placement.
The design is shown in Fig.~\ref{cap_base_fibre}.
The package is of low cost and the cap with the fibers attached can be
mounted to the base near the end of the pixel detector assembly.

In the OSU design, precise alignment of the fibres to the VCSELs is
achieved by fabricating the bases and caps with high precision and
by the accurate placement of the VCSELs and fibres relative to the
base and cap, respectively.
The precision dice placement is realized using a jig with precisely
machined pockets for the base, VCSELs and PIN.
The accurate placement of the fibres in the cap is achieved by
precise machining of the holes in the injection mold used to
fabricate the cap (see below).

\section{Prototype Results}
\vspace*{-0.5pt}
\noindent
We have investigated several kinds of material for the fabrication
of the package.
The first material tested was aluminum silicate, a machinable ceramic.
We then decided to switch to macor because of its smaller grain size,
which was better in reproducing some of the fine features in the design.
Unfortunately, the adhesion of the tiny VCSEL to the gold
trace was not good because the gold surface was not smooth
enough due to the roughness of the machined macor surface.
We now use aluminia to fabricate the base because of its
much smoother surface.
To fabricate the bases, aluminia sheet is ground to the precise
thickness of the bases and then cut into strips for deposition
of three-dimensional traces.\cite{HT}
Most of the deposited traces have good connectivity across
the corner of the bases.
Strips with a large number of traces of good connectivity
were then precisely diced into individual bases.\cite{AD}

We have also attempted to fabricate the cap using the same
materials, aluminum silicate and macor.
However, it is difficult to consistently machine the cap to the desired
tolerances.
We now use Ultem (polyetherimide), a mold-injectable plastic
with a radiation tolerance\cite{rad} of 10 GRad.
Due to the high cost and long lead time in developing the mold
injection technique for fabricating small parts with high precision,
we decided to use ``manual mold injection''.
Here a small spring-loaded mold of 5~cm $\times$ 5~cm $\times$ 10~cm
is used in a small oven as opposed to a standard mold of
30~cm $\times$ 30~cm $\times$ 30~cm placed inside an automatic mold
injection machine of 1~m $\times$ 1~m $\times$ 2~m.
The critical part in the mold fabrication is the precisely machined
mold for the interior of the cap that goes over the base and the
three precisely located holes for the placement of the three pins
which produce the holes for the fibers.
We have proven the principle of this precise micro-mold injection
technique and can fabricate several quality caps per hour.

We have fabricated twelve optical packages.
The coupled power of ten of the packages are above the specification
of 300~$\mu$W minium power for an average current of 10~mA in the VCSEL.
The caps are interchangeable, an important test of the feasibility
of the connector-type design.
The waveforms\cite{Oxford} in both VCSELs have fast rise time, $<$ 1ns.
The PIN diode also has good responsivity, 0.5~A/W.
The bit error rate (BER) meets the specification, $<$ 10$^{-9}$ at 90\%
confidence level, for PIN current $>$ 60~$\mu$A.
The cross talk between the PIN diode and the inner VCSEL can produce a data
bit error and this can be measured by sending the signal from the outer
VCSEL to the PIN diode while running the inner VCSEL with an asynchronous
40~MHz clock at an average current of 10~mA.
There is no measured cross talk (BER $<$ 10$^{-9}$ at 90\% C.L.)
between the inner VCSEL and the PIN diode for PIN current above $\sim 20~\mu$A.
The ground plane around the traces is critical in achieving this low cross talk.

\section{Summary}
\vspace*{-0.5pt}
\noindent
In conclusion, we have demonstrated the principle of the connector-type
optical package.
The package can be fabricated at low cost and satisfies the
minimum coupled power and low cross talk requirements.

\nonumsection{Acknowledgements}
\noindent
This work was supported in part by the U.S.~Department of Energy.
The author wishes to thank K. Arms, J. Burns, H.P. Kagan, R.D.~Kass, S. Smith,
T. Weidberg, and R. Wells for their contributions.

\nonumsection{References}

\end{document}